\documentclass[%
 aip,
 jmp,%
 amsmath,amssymb,
 reprint,%
 nofootinbib,
 groupedaddress,
]{revtex4-2}
\usepackage{booktabs}
\usepackage{enumitem}
\usepackage[table]{xcolor}
\usepackage{graphicx}
\usepackage{dcolumn}
\usepackage{bm}
\usepackage{subfigure}

\begin{document}

\preprint{AIP/123-QED}

\title[Impact of Remote Learning on Students' Ability to Stay on Track]{Analyzing the Heterogeneous Impact of Remote Learning on Students' Ability to Stay on Track During the Pandemic}

\author{Zhongzhou Chen}
 \email{Zhonzhou.Chen@ucf.edu}
\author{Tom Zhang}%
 \email{Tom.Zhang@ucf.edu}
\affiliation{ 
Department of Physics, University of Central Florida, 4000 Central Florida Blvd, Orlando, Florida, 32816
}%

\begin{abstract}
This study investigates how remote learning due to the COVID pandemic impacts students' ability to keep up with the pace of instruction in a university level physics course, with a focus on the heterogeneous impact of remote learning on different demographic groups. Student learning data is collected from 70 online learning modules assigned as both online homework and self-study material in both Fall 2020 and Spring 2020 semesters, with the first 41 modules being assigned before campus closure in Spring 2020. Students' ability to stay on track is measured by three data indicators: percentage of modules submitted before the due date, percentage of modules submitted early for extra credit, and percentage of modules properly engaged with. The student population is divided into two demographic groups according to each of the four demographic variables: Gender, Ethnicity, Transfer Status and First generation status. All three data indicators from both semesters are first compared between each pair of demographic groups within the same semester, and then compared within the same group between the two semesters. Exam and course scores are also compared between demographics groups in Fall 2020 and against scores from the Fall 2019 semester. The analysis revealed that remote learning significantly reduced on-time and early submission for first-generation students, but increased the frequency of early submission for transfer students. Under-represented minority students had higher percentage of engaged modules during the pandemic, while non-URM students had the same level of module engagement before and after the pandemic. These results suggest that pandemic remote learning had very different impact for students with different demographic background, and future instructional design should strive to provide more flexible options to meet the various needs for a diverse higher-ed student population.
\end{abstract}

\maketitle

\section{Introduction}
More than a year after the COVID-19 pandemic forced educational institutions around the world to abruptly switch to remote learning, there is now a growing body of research studying the impact of the pandemic on various aspects of student learning. While many studies found significant negative impacts on student performance \cite{Biwer2021, Orlov2021}, engagement and goal orientation \cite{Daniels2021}, mental health \cite{Elmer2020}, and time of graduation \cite{Aucejo2020}, others found little or no overall impact \cite{ElSaid2021} or even significant positive impacts \cite{Gonzalez2020}. Regarding physics courses, Wilcox and Vignal \cite{Wilcox2020} reported mixed student perception of remote learning, and lack of study space and campus resource as the most prominent challenges. Fox et al.\cite{Fox2021} found no differences in overall E-CLASS scores, with two questions showing a significant pre-post shift towards the more expert-like direction. Klein et al. \cite{Klein2021} found an overall positive attitude towards distance learning, but also reported that students faced difficulties keeping up with the usual pace of learning. 

Notably, a number of studies have revealed that the impact of the pandemic is highly heterogeneous: while some groups of students faced substantial hurdles, others experienced little to no impact or even benefited from remote learning. For example, Biwer et al. \cite{Biwer2021} identified four clusters of students for which they referred to as  "the overwhelmed", "the adapters", "the maintainers", and "the surrenderers". Aucejo et al. \cite{Aucejo2020} found that students from disadvantaged groups, in particular non-white, first generation and low-income students are significantly more likely to delay graduation, change majors, or have lower overall GPA. Gillis and Krull \cite{Gillis2020} also reported that minority students had a more negative experience during the pandemic. Naujoks et al. \cite{Naujoks2021} reported significant variation in students' ability to manage resources and time, with females being significantly more likely to employ effective resource management strategies.

The vast majority of studies on the impact of pandemic remote learning are based on students' survey responses. While survey results provide insights into a wide variety of student perceptions and experiences, its validity could be affected by low response rates and self-selection effects \cite{Klein2021}. Moreover, as Naujoks et al. \cite{Naujoks2021} pointed out, using online questionnaires might have limited study participation from students with less stable internet connections, which are also likely those who suffered the most negative impacts of remote learning. A number of studies also compared summative assessment data and course GPA between face-to-face (F2F) and remote learning situations. However, the condition under which exams were designed, administered, and scored were in many cases very different between F2F and remote classes \cite{Wilcox2020} which raises concerns regarding the validity of such comparisons.

More importantly, summative assessment scores or questionnaires can be less sensitive to shorter term impacts of pandemic remote learning. For example, a student who was temporarily lagging behind schedule due to COVID related challenges could catch up afterwards by exerting extra effort and eventually obtain satisfactory assessment scores and course outcome. In fact, some studies suggested that a major difficulty faced by many students during the pandemic is trying to keep up with the usual pace of learning, i.e. "staying on track" \cite{Klein2021, Chhetri2020}.

So far, very few published studies have utilized data from learning management systems (LMS) and online learning platforms to gain direct insight into students' learning behavior during the pandemic remote instruction period, especially their ability to stay on track. In an earlier paper \cite{Zhang2021}, we extracted six types of learning behaviors indicative of students' self-regulated learning processes from students' online log data collected during the emergency remote learning period of April 2020. We compared the frequency of those behaviors before and after the abrupt shift to remote learning and found no differences in outcome measures such as number of homework assignments completed on time, but a significant increase in guessing and skipping of learning resources. 

The current study utilizes online log data collected from students enrolled in a large (>250 students) calculus based introductory physics course taught remotely during the Fall 2020 (FA20) semester, as well as data from the same course taught face to face during the first 11 weeks of the Spring 2020 (SP20) semester. The current analysis seeks to reveal the impact of remote instruction on students' ability to stay on track of the course, focusing in particular on whether and how different demographic groups are impacted heterogeneously by remote instruction. More specifically, we will seek to answer the following research questions:

\begin{enumerate}[label=RQ\arabic*:]
    \item How did pandemic remote learning impact students' ability to keep up with the pace of instruction i.e., how often did students ``fall off track" during the pandemic?
    \item Did pandemic remote learning have different impacts on different demographic groups?
\end{enumerate}

The data we use to answer those questions were collected from students' interaction with 70 online learning modules (OLMs) \cite{Chen2020} assigned as both homework and learning materials throughout the semester. In FA20, all modules were administered as the course was taught in remote mode while in SP20, the first 41 modules were administered while the course was taught in F2F mode. 

To answer RQ1, we measured students' ability to "stay on track" by three data indicators: 

\begin{enumerate}
    \item the fraction of OLMs completed before the assignment due date
    \item the fraction of OLMs completed earlier than the due date for extra credit
    \item the fraction of OLMs that were properly engaged rather than skimmed through
\end{enumerate}

The detailed definition of each indicator is described in section \ref{sec: methods}.

To answer RQ2, we divided the student population into two groups along each of the four demographic variables: gender, race/ethnicity, enrollment status, and first-generation status. We then compared the three data indicators between each of the four pairs of demographic groups. In addition, we also compared summative exam scores and course total scores between demographic groups in FA20, and between FA20 and Fall 2019 (FA19) semesters. Summative scores from SP20 were not used as they were obtained under a mixture of F2F and emergency remote instructional modes. 

We will explain and justify in detail the instructional conditions, data definition, and the data analysis and comparison schemes in the next section (section \ref{sec: methods}). In section \ref{sec: results} we will present the results of the analysis, followed by a discussion of the implications and limitations in section \ref{sec: discussion}.

\section{Methods}\label{sec: methods}
\subsection{Instructional Conditions}
\subsubsection{General Information}

Data for the current analysis were collected from a university level calculus based introductory physics I course taught at the University of Central Florida during the FA19, SP20, and FA20 semesters. 

In FA20, the entire course was taught remotely with 70 OLMs assigned as homework. Two mid-term exams and one final exam were all administered online without human proctor but with students' webcam taking frequent snapshots. Bi-weekly quizzes were administered during recitation sessions, proctored by a TA in a hybrid setting. Content delivery was achieved through a combination of OLMs, OpenStax textook, and pre-recorded instructor lecture videos. Students worked in groups on problem solving worksheets during optional synchronous class meetings, and interacted with the instructor through Microsoft Teams. \cite{Chen2021}

In SP20, the first 10 weeks of the course were taught in F2F modality and the last 5 weeks in an emergency remote mode using Microsoft Teams following COVID induced campus shutdown. The same 70 OLMs were assigned as homework assignments, with the first 41 modules assigned during the first 10 weeks prior to campus closure. Only the first mid-term exam was administered in a proctored F2F environment. The second mid-term and final exam were administered online with webcam snapshots. No quizzes were conducted throughout the semester. Students were required to attend synchronous lectures during the F2F period, and learned from asynchronous lecture videos during the remote period.

In FA19, the entire course was taught in F2F mode with synchronous lecturing. Homework assignments consisted of a mixture of 44 OLMs and problems from a commercial online homework platform.  Two mid-term exams and one final exam were administered on paper in a proctored environment, and no quizzes were administered.

\subsubsection{OLM Online Homework}
An OLM is a form of online instructional design that integrates assessment, practice, and instruction into a single module focusing on one unit concept e.g., the definition of kinetic energy \cite{Chen2020, Chen2018, Chen2018a}. Students have a total of five attempts on the assessment of each OLM, which contains 1 or 2 problems, and must make one initial attempt to be able to access the instructional materials of the module. A typical OLM sequence contains 5-12 OLMs that cover one topic (e.g., conservation of mechanical energy) assigned over a 1-2 week period and due on the same day. Students must complete each OLM in the same sequence in the given order. Completion is defined as either passing the assessment on any given attempt or using up all 5 attempts. Accessing the instructional material is not required for completing the OLM, although the instructional materials are designed specifically to help students answer the problems. In both SP20 and FA20 semesters, students could complete OLMs past the due date but would receive a 13\% daily penalty for each late module. Interested readers can experience the OLMs as a student following the URL at: \url{https://canvas.instructure.com/courses/1726856} 
\subsubsection{Extra Credit for Early Submissions}

In both SP20 and FA20 semesters, students could receive a small amount of extra course credit for completing some of the OLMs at earlier dates than the due date of the entire sequence. For example, in a 10 module sequence assigned over 2 weeks, students will have three opportunities for extra credit: submitting OLMs 1-3 ten days ahead, OLMs 4-6 seven days ahead, and OLMs 7-9 3 days ahead of the sequence due date. As described in Felker et al. \cite{Felker2020}, this procedure significantly promotes better work distribution. In FA20, 59 of the 70 OLMs (30 out of the first 41 modules) were available for early submission credits. Due to small scheduling differences in SP20, 33 out of the first 41 modules were available for extra credit. 
\subsubsection{Student Demographics}
The following four types of demographic information is obtained for each student from the institutional knowledge management office at UCF.
\begin{enumerate}
    \item Gender\footnote{This may not necessarily reflect the gender identity of the student in all cases}: UCF's institutional records categorizes each student as either Male or Female.
    \item Race/Ethnicity: To simplify the current analysis, we classify students according to whether or not they identify as under-represented minorities (URMs) in STEM disciplines which consists of American Indians, Alaskan Natives, Black/African Americans, Hispanic/Latino, and Pacific Islanders.
    \item Transfer Students: A student is classified as a transfer student if they previously attended one or more higher educational institutions and completed 12 or more transferable college or university credits post-high school. At UCF, the vast majority of transfer students come from local two-year state colleges.
    \item First Generation College Students: A student is classified as a first generation college student if neither their parents nor guardians have a bachelor's degree.
\end{enumerate}
\subsection{Data Indicators for "Staying on Track"}
The following indicators were selected to reflect students' ability to keep up with the pace of instruction throughout the semester i.e., being able to stay on track.

\begin{enumerate}
    \item \textbf{On-Time Submission}: The fraction of OLMs completed prior to the due date of each sequence, including those submitted before the early submission dates. 
    \item \textbf{Early Submission}: The fraction of available OLMs completed prior to the respective early submission dates.
    \item \textbf{Engaged Modules}: A student is classified as having properly engaged with an OLM if one of the two behaviors is observed:
    \begin{enumerate}
        \item Passing the assessment on the first or second attempt without accessing the instructional material, with the passing attempt being longer than 35 seconds. 
        \item Studied the instructional materials in the module after the first or second failed attempt and either passed the assessment on subsequent attempts or used up all attempts.

    \end{enumerate}
            For a student to be considered as having studied the instructional material, the student must both spend longer than 35 seconds on the instructional materials and view all the practice questions embedded in the instructional material
\end{enumerate}

In two earlier studies on OLMs \cite{Chen2020, Guthrie2020}, those types of engaged behaviors are observed to be more frequently observed among higher performing students. Behaivor (a) was indicative of higher incoming knowledge, whereas behavior (b) was often associated with students who needs to learn the material, and more often observed on harder modules. Since we are only measuring level of engagement, not level of mastery, we include in behavior (b) students who both passed and failed the assessment after having studied the instructional materials. 

On the other hand, other types of behaviors such as passing on first attempt under 35 seconds, or making 3 or more consecutive attempts without accessing the instructional materials are more frequently observed among lower performing students and on more challenging OLMs, as well as during the abrupt remote learning period in the last 5 weeks of the SP20 semester \cite{Zhang2021}.

Under the current definition, the fraction of engaged module is significantly positively correlated with average quiz score (Pearson's $r^2 = 0.2051, p < 0.01$ t-test) for most students in the FA20 semester. On the other hand, more stringent criteria for studying the instructional materials (e.g., answering or viewing the solutions to all practice questions) did not significantly improve the correlation. As a result, we did not include a separate "highly engaged" criteria as another data indicator. Here we chose average quiz score as a measure of students' learning outcome since quizzes were administered more frequently in a proctored setting, which makes them better indicators of students' mastery of skills than exams during pandemic remote instruction.

\subsection{Data Analysis Scheme}
\subsubsection{Comparing the Frequency of Data Indicators}
For each of the three data indicators mentioned above, we first compared their observed frequency between each pair of demographic groups divided according to the four demographic variables. Mann-Whitney U tests were employed to test the significance of each comparison. The analysis was conducted using all of the OLMs assigned in the FA20 semester, the first 41 OLMs assigned in the FA20 semester, and the same 41 OLMs assigned in the SP20 semester during the F2F instructional period. Data from FA19 was not included in the above comparisons since OLMs were mixed with other forms of online homework and the homework due date policy changed between FA19 and SP20. 

If a significant gap in a data indicator is observed between two demographic groups in one semester but not the other, it would suggest that a potential impact could be due to pandemic remote instruction. For example, if first generation students were observed to have significantly lower fractions of early submissions in FA20 compared to non-first generation students but the same difference was not observed in the SP20 semester, then its is likely that pandemic remote instruction impacted early submission frequencies of one or both groups.

In those cases, we then compare the three indicators for the same demographic group between FA20 and SP20 semesters to determine:

\begin{enumerate}
    \item Whether the impact of the pandemic remote learning was concentrated on on of the two groups, or could be observed on both.
    \item Whether the impact is in the positive or negative direction.
\end{enumerate}

For example, did first generation students make less early submissions in FA20 than in SP20, or did non-first generation students make more early submissions in FA20, or both? The statistical significance of all comparisons are obtained by Mann-Whitney U test, whereas the magnitude and direction of any significant impact is presented visually as described below.

\subsubsection{Visualization of the Distribution of Data Indicators}
\begin{figure}[h]
    \centering
    \includegraphics[width = .8\textwidth]{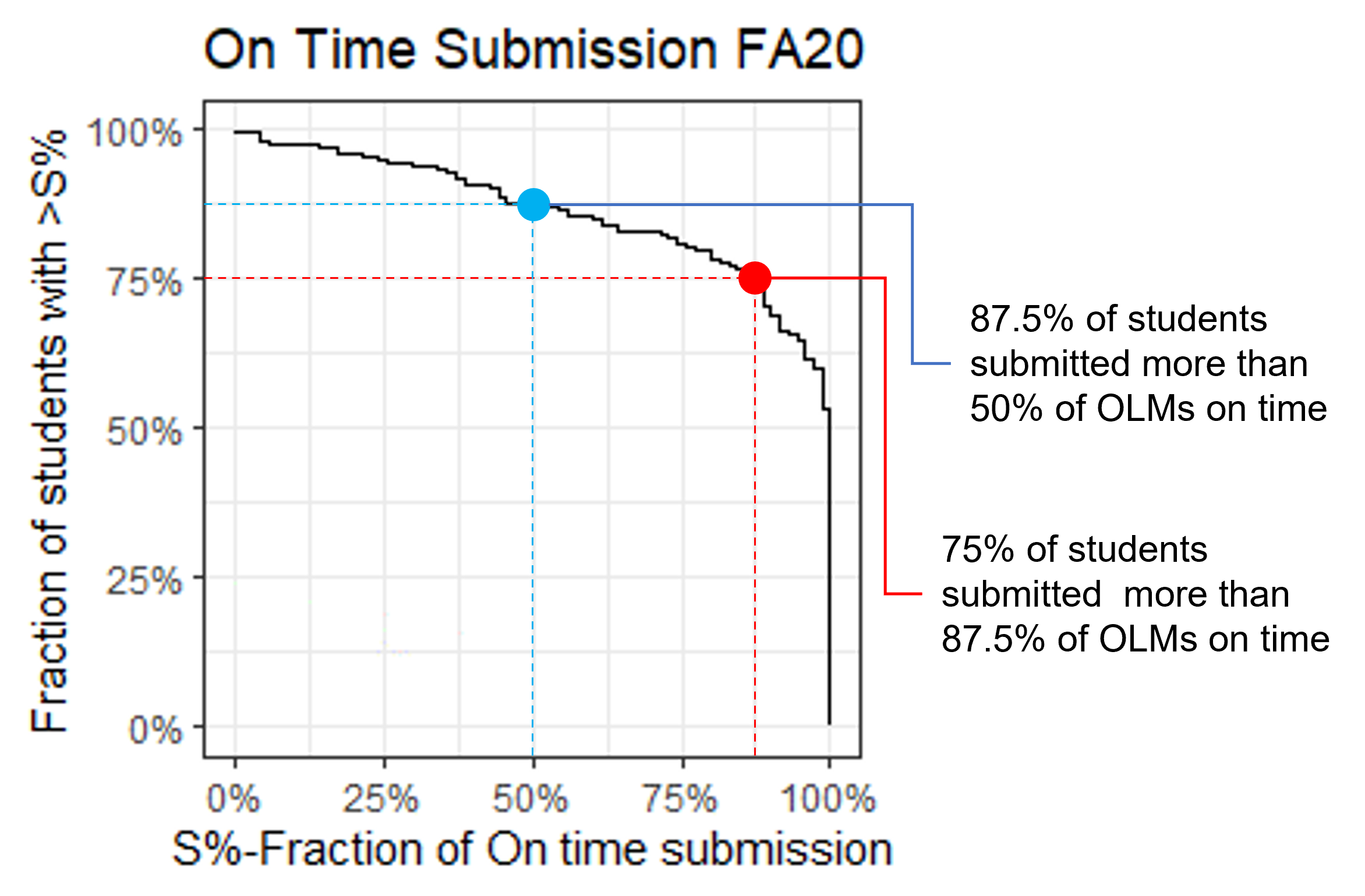}
    \caption{Complementary Cumulative Distribution (CCD) graph of the distribution of on-time submission fraction for all students in FA2020 semester.}
    \label{fig: CCD}
\end{figure}

We visualized both the overall distribution of each data indicator and any significant gaps between two groups using complementary cumulative distribution (CCD) graphs following the precedence of two earlier studies \cite{Seaton2014, Seaton2014_2}. We chose to use CCD graphs because the frequency distributions can be highly non-normal, especially for smaller demographic groups. As illustrated in Figure \ref{fig: CCD}, a point on a zig-zag line in a CCD graph represents the fraction of students (y-coorinate) that has greater than a certain fraction (x-coordinate) of a given indicator, such as on-time submission.

In all CCDs comparing a data indicator between two demographic groups, the minority group is represented with a red line and the majority group with a cyan line. When comparing the same demographic group across two semesters, data from FA20 is represented with a red curve and that of SP20 with a cyan curve. Only the CCDs of statistically significant comparisons are shown.

\subsubsection{Multivariate Linear Models}
We also created logistic multi-variate linear models to systematically investigate whether impacts of pandemic remote learning was more significant among sub-groups with multiple demographic labels, such as female-URM students. The frequencies of each data indicator are used as responses. Since their values are bounded and their distributions are non-normal, the frequency data underwent logistic transformation. The covariates include all four demographic labels as well as all possible interaction terms between the labels.

\subsubsection{Comparing Summative Course Scores}
Students' normalized summative scores including average quiz scores, average exam scores, and total course scores in FA20 are also compared between majority and minority groups based on all for demographic dimensions. Normalized exam scores and total course scores are also compared to data from FA19 following the same analysis scheme above. Grades for SP20 semester were not used for this comparison as SP20 was taught with a mixture of F2F and remote modes, furthermore the remote instruction was organized differently from FA20. It must be emphasized that there existed significant differences in exam conditions and course grading schemes between the FA19 and FA20 semesters, thus the outcomes of this comparison should be interpreted with caution. Statistical significance of the grade comparisons are also obtained sing Mann-Whitney U tests.

\subsubsection{Technical Details}
The OLMs are implemented on the Obojobo Learning Objects platform \cite{CenterforDistributedLearning}, developed by the Center for Distributed Learning at UCF. Statistical analysis and visualization of students' online learning behavior is conducted using R \cite{R} and the tidyverse package \cite{Wickham2019}. 

\section{Results}\label{sec: results}
\subsection{Student Population}
251, 273, and 268 students enrolled in the course and completed one exam during the FA20, SP20, and FA19 semesters respectively. The breakdown of students according to each of the four demographic dimensions are listed in Table \ref{tab: demOverall}, in which the ``Min." column listed the size of the minority group of each of the demographic variables: Female, First-Generation, URM, and Transfer students. The SP20 class had significantly less transfer students ($p < 0.01$, Fisher's exact test), while the FA19 class had fewer female students ($p < 0.03$, Fisher's exact test). 
\begin{table}[h!]
    \caption{Number of students in the minority and majority groups of each demographic variable in all three semesters.}\label{tab: demOverall}
    \centering
    \begin{tabular}{c c c c c c c}
        \toprule\toprule
         & \multicolumn{2}{c}{FA2020} & \multicolumn{2}{c}{SP2020} & \multicolumn{2}{c}{FA2019}\\
         \arrayrulecolor{black!30}\midrule 
         Dimension & Min. & Maj. & Min. & Maj. & Min. & Maj. \\
         \arrayrulecolor{black}\midrule 
         Gender & 67 & 184 & 76 & 197 & 52 & 217 \\
         \arrayrulecolor{black!30}\midrule 
         Frst.Gen & 46 & 205 & 43 & 230 & 56 & 212 \\
         \midrule
         URM & 98 & 153 & 108 & 165 & 96 & 172\\
         \midrule
         Transfer & 70 & 181 & 47 & 226 & 78 & 190\\
         \arrayrulecolor{black}\bottomrule\bottomrule
    \end{tabular}
\end{table}

\subsection{Comparing Data Indicators}
\subsubsection{On-Time Submissions}
Figure \ref{fig: CCD} shows the distribution of on-time submission fractions for all students in the FA20 semester. As shown in the figure, more than half of the students submitted all OLMs on time, and 75\% of students submitted at least 87.5\% of OLMs on time.

\begin{figure}[h!]
    \centering
    \includegraphics[width=.8\textwidth]{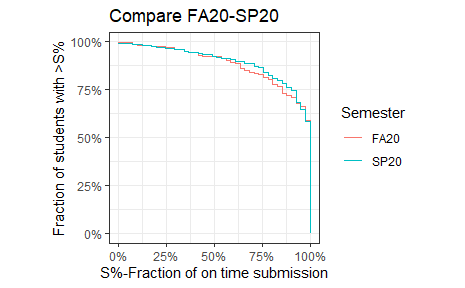}
    \caption{CCD of the fraction of on-time submission of the first 41 modules, compared between FA20 and SP20 semesters.}
    \label{fig: onTime_FASP}
\end{figure}

For the first 41 modules, there were no statistically significant differences in the fraction of OLMs submitted on time between FA20 and SP20. As shown in Figure \ref{fig: onTime_FASP}, in both semesters, more than 50\% of students submitted all OLMs on time and about 75\% of students submitted more than 90\% of the OLMs on time.

\begin{figure}[!h]
    \includegraphics[width=.8\textwidth]{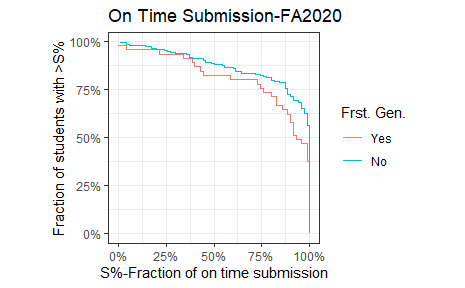}
    \caption{CCD of fraction of on-time submission compared between first generation and non-first generation students in FA2020}
    \label{fig: onTime_firstWithin_FA20}
\end{figure}

Table \ref{tab: onTime_within} shows the statistical test results of comparing the distribution of on-time submissions between minority and majority groups according to each of the four demographic variables. In FA20, first generation students were found to have submitted a significantly lower fraction of modules on-time compared to non-first generation students. The difference is still significant for the first 41 modules in FA20, but not for the same OLMs in SP20. Figure \ref{fig: onTime_firstWithin_FA20} shows the gap in on-time submissions between first generation and non-first generation students in FA20. As shown in the figure, while 75\% of non-first generation students submitted more than 87\% of OLMs on time, only about 65\% of first generation students were able to do the same.

\begin{table}[h!]
    \caption{Statistical test results comparing the frequency of on-time submission according to demographic division. FA2020-Pre and SP2020-Pre contain results from the first 41 modules, assigned pre-COVID in SP2020.}\label{tab: onTime_within}
    \centering
    \begin{tabular}{c c c c c c c}
        \toprule\toprule
         FA2020 & \multicolumn{2}{c}{FA2020} & \multicolumn{2}{c}{FA2020-Pre} & \multicolumn{2}{c}{SP2020-Pre} \\
         \arrayrulecolor{black!30}\midrule 
         Dimension & Statistic & p & Statistic & p & Statistic & p \\
         \arrayrulecolor{black}\midrule 
         Gender & 5261 & 0.07 & 5461 & 0.17 & 7050 & 0.47 \\
         \arrayrulecolor{black!30}\midrule 
         Frst.Gen & \textbf{3740} & \textbf{0.03} & \textbf{3736} & \textbf{0.03} & 4776 & 0.69 \\
         \midrule
         URM & 7292 & 0.73 & 7035 & 0.44 & 9946 & 0.07 \\
         \midrule
         Transfer & 6959 & 0.12 & 6816 & 0.17 & 4604 & 0.11 \\
         \arrayrulecolor{black}\bottomrule\bottomrule
    \end{tabular}
\end{table}

However, between semester comparisons for both first generation and non-first generation students did not detect any statistically significant differences, as shown in the "On-Time" column of Table \ref{tab: between}, nor are there any other differences detected for other demographic groups.

\begin{table}[h!]
    \caption{Results of all comparisons within the same demographic group and between FA20 and SP20 semesters for the first 41 modules. Statistically significant results at $\alpha < 0.05$ level are highlighted in bold.}
    \label{tab: between}
    \centering
    \begin{tabular}{c c c c c c c c}
        \toprule\toprule
        & & \multicolumn{2}{c}{on-time} & \multicolumn{2}{c}{Early} & \multicolumn{2}{c}{Engage} \\
         \midrule
         Dimension & Type & Statistic & p & Statistic & p & Statistic & p \\
         \midrule 
         Gender & Male & 17771 & 0.65 & 19177 & 0.37 & 18506 & 0.09 \\
         \arrayrulecolor{black!30}\midrule 
         Gender & Female & 2554 & 0.71 & 2653 & 0.46 & 2546 & 0.23 \\
         \midrule 
         Frst.Gen. & Yes & 798 & 0.12 & 742 & 0.06 & 870 & 0.27 \\
         \midrule 
         Frst.Gen. & No & 24327 & 0.52 & \textbf{26383} & \textbf{0.031} & 24608 & 0.08 \\
         \midrule 
         URM & Yes & 5358 & 0.76 & 5326 & 0.84 & 4228 & 0.46 \\
         \midrule 
         URM & No & 12240 & 0.59 & 13737 & 0.17 & \textbf{14678} & \textbf{0.0014} \\
         \midrule 
         Transfer & Yes & 1659 & 0.71 & \textbf{1997} & \textbf{0.02} & 1289 & 0.96 \\
         \midrule 
         Transfer & No & 20519 & 0.96 & 21056 & 0.68 & \textbf{21562} & \textbf{0.033} \\
         \arrayrulecolor{black}\bottomrule\bottomrule
    \end{tabular}
\end{table}

The multivariate linear model did not identify any intersectional groups that are significantly correlated with the fraction of on-time submissions in the FA20 semester. For the SP20 semester, the linear model identified two interaction terms significantly correlated with on-time submission frequency. However, the two groups contain 1 and 7 students respectively, too few to make any meaningful conclusions.

\subsubsection{Early Submissions}
\begin{figure}[h!]
    \noindent
    \centering
    \makebox[\textwidth]{
        \subfigure[]{
            \includegraphics[width=.57\textwidth]{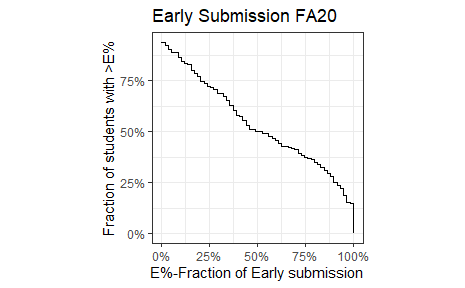}
            \label{fig: early_FA20}
        }
        \subfigure[]{
            \includegraphics[width=.57\textwidth]{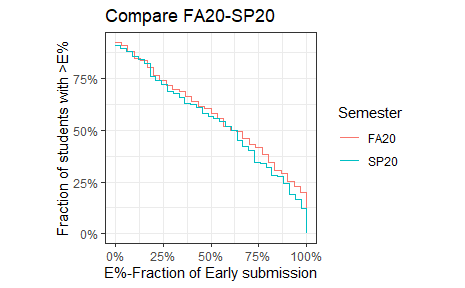}
            \label{fig: early_FASP20}
        }
    }
    \caption{CCD of fraction of early submission: (a)for all modules in FA20 semester. (b) for the first 41 modules in FA20 and SP20 semesters.}
\end{figure}

For early submissions, 50\% of students in FA20 submitted more than 40\% of the OLMs prior to the early submission date (Figure \ref{fig: early_FA20}). No statistically significant differences were detected between SP20 and FA20 semesters regarding the fraction of early submissions for the first 41 modules (Figure \ref{fig: early_FASP20}).

\begin{figure}[h!]
    \noindent
    \centering
    \makebox[\textwidth]{
        \subfigure[]{
            \includegraphics[width=.53\textwidth]{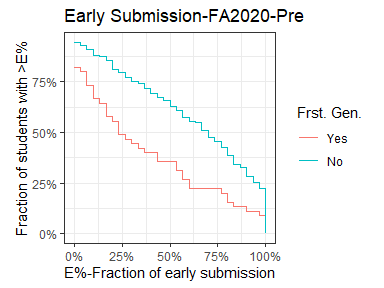}
            \label{fig: early_firstWithin}
        }
        \subfigure[]{
            \includegraphics[width=.53\textwidth]{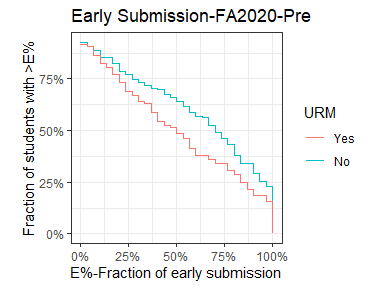}
            \label{fig: early_URMWithin}
        }
    }
\end{figure}

Comparison between demographic groups revealed that first generation students in FA20 had significantly lower fraction of early submission compared to non-first generation students, as listed in Table \ref{tab: early_within}. URM students also had slightly lower fraction of early submissions but the difference was not significant at the $\alpha = 0.05$ level. During the first 41 OLMs in FA20, both first generation and URM students had significantly lower early submission frequencies but this difference was not present for the first 41 modules of SP20. As shown in Figure \ref{fig: early_firstWithin}, the gap among first generation students is particularly wide: while 50\% of non-first generation students submitted more than 50\% of the modules early for the entire semester, only about 25\% of first generation students were able to do the same. In comparison, the gap between URM and non-URM populations is narrower. During the F2F period of the SP20 semester, transfer students had significantly lower fraction of early submissions compared to non-transfer students, yet the same difference is not observed between transfer and non-transfer students in FA20. 

\begin{table}[h!]
    \caption{Statistical test results comparing fraction of modules submitted early between different demographics groups within the same semester.}
    \label{tab: early_within}
    \centering
    \begin{tabular}{c c c c c c c}
        \toprule\toprule
         FA2020 & \multicolumn{2}{c}{FA2020} & \multicolumn{2}{c}{FA2020-A} & \multicolumn{2}{c}{SP2020-A} \\
         \arrayrulecolor{black!30}\midrule 
         Dimension & Statistic & p & Statistic & p & Statistic & p \\
         \arrayrulecolor{black}\midrule 
         Gender & 5278 & 0.10 & 5560 & 0.31 & 6937 & 0.40 \\
         \arrayrulecolor{black!30}\midrule 
         Frst.Gen & \textbf{3244} & \textbf{0.0016} & \textbf{2778} &  \textbf{0.00} & 5380 & 0.36 \\
         \midrule
         URM & 6465 & 0.07 & \textbf{6200} & \textbf{0.03} & 9816 & 0.16 \\
         \midrule
         Transfer & 6174 & 0.93 & 6371 & 0.72 & \textbf{3952} & \textbf{0.01} \\
         \arrayrulecolor{black}\bottomrule\bottomrule
    \end{tabular}
\end{table}

\begin{figure}[h!]
    \noindent
    \centering
    \makebox[\textwidth]{
        \subfigure[]{
            \includegraphics[width=.53\textwidth]{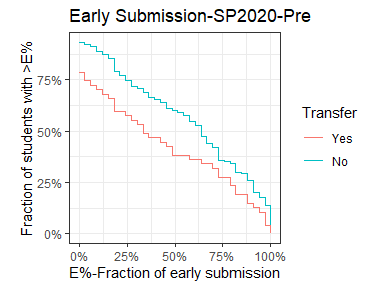}
            \label{fig: early_transferWithin}
        }
    }
    \caption{CCD of fraction of early submission where significant differences between minority and majority demographic groups are detected: (a) Comparison between First Gen. and Non-First Gen. students in FA20 (b) Comparison between URM and non-URM students in FA20. (c) Comparison between Transfer and non-Transfer students in SP20.}
\end{figure}

\begin{figure}[h!]
    \centering
    \noindent
    \makebox[\textwidth]{
        \subfigure[]{
            \includegraphics[width=.53\textwidth]{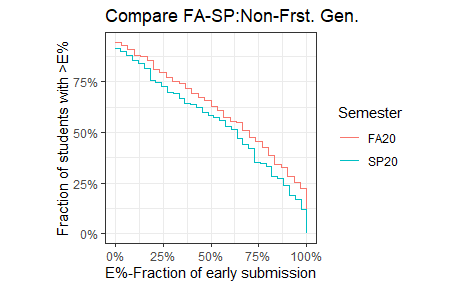}
            \label{fig: early_between_nonfirst}
        }
        \subfigure[]{
            \includegraphics[width=.53\textwidth]{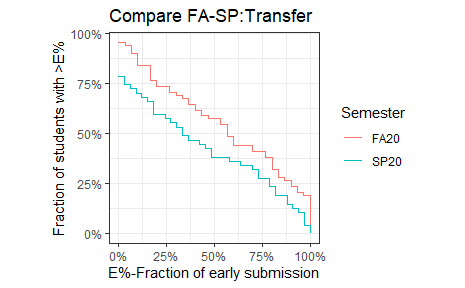}
            \label{fig: early_between_transfer}
        } 
    }
    \caption{CCD of fraction of early submission compared within the same demographic group between FA20 and SP20 semesters: (a) Comparison for non-Frst.Gen students. (b) Comparison for transfer students.}
    \label{fig: early_between_semesters}
\end{figure}

As shown in the ``Early" column of Table \ref{tab: between} and in Figure \ref{fig: early_between_semesters}, between semester comparisons revealed that first generation students had lower early submission frequencies during FA20 when compared to SP20 , while non-first generation students had significantly higher early submission frequencies (Figure \ref{fig: early_between_nonfirst}). The magnitude of the gap is greater for first generation students, but the p-value is only 0.06 likely due to the smaller sample size. On the other hand, transfer students had significantly higher early submission frequencies during FA20 compared to SP20 (Figure \ref{fig: early_between_transfer}), while no significant difference is detected for non-transfer students. 

The multivariate linear model did not identify any groups with multiple demographic labels that are significantly correlated with the fraction of early submissions in either semester.

\subsubsection{Engagement with the OLMs}
\begin{figure}[h!]
    \noindent
    \centering
    \makebox[\textwidth]{
        \subfigure[]{
            \includegraphics[width=.57\textwidth]{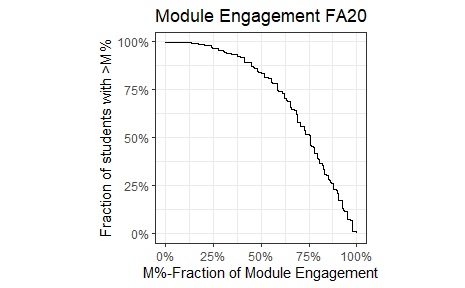}
            \label{fig: engage_FA20}
        }
        \subfigure[]{
            \includegraphics[width=.57\textwidth]{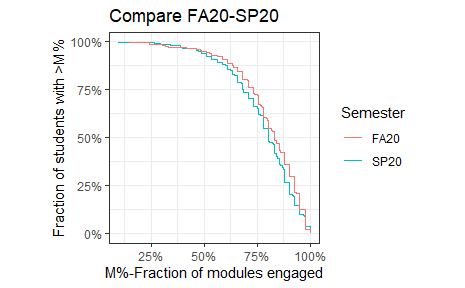}
            \label{fig: engage_FASP20}
        }
    }
    \caption{CCD of fraction of engaged modules: (a) all modules in FA20 semester (b) the first 41 modules compared between FA20 and SP20 semesters.}
\end{figure}

As shown in Figure \ref{fig: engage_FA20}, half of the students in FA20 engaged with more than 75\% of all the OLMs.  

For the first 41 modules, FA20 students somewhat unexpectedly had a higher overall fraction of engaged modules compared to SP20.  As shown in Figure \ref{fig: engage_FASP20}, the difference is only a few percent across the board but still statistically significant ($U = 34806,\, p = 0.038$). To examine whether the difference could be caused by either higher incoming knowledge or more students obtaining answers from other sources, we did the comparison again while excluding those who passed each module within 2 attempts before accessing the instructional materials, and the gap remained statistically significant.

\begin{figure}[h!]
    \centering
    \includegraphics[width=.8\textwidth]{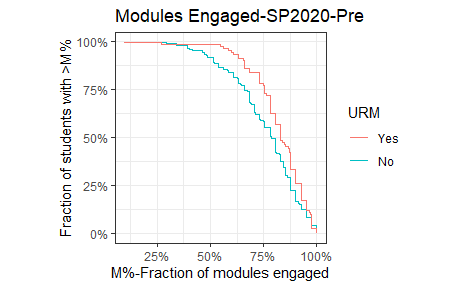}
    \caption{CCD of fraction of engaged modules for URM and non-URM students for the first 41 modules in SP2020 semester.}
    \label{fig: engage_URMWithin}
\end{figure}

When comparing between demographics groups, none of the four comparisons resulted in significant differences for the FA20 semester either for the entire semester or for the first 41 modules (Table \ref{tab: engage_within}). On the other hand, URM students were significantly more engaged compared to non-URM students in SP 2020 (see ``Engagement" column of Table \ref{tab: between}, and Figure \ref{fig: engage_URMWithin}). 

\begin{figure}[b]
    \noindent
    \centering
    \makebox[\textwidth]{
        \subfigure[]{
            \includegraphics[width=.53\textwidth]{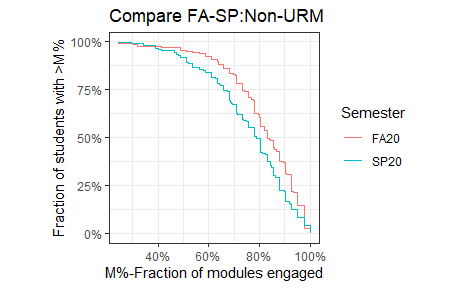}
            \label{fig: engage_betweenNURM}
        }
        \subfigure[]{
            \includegraphics[width=.53\textwidth]{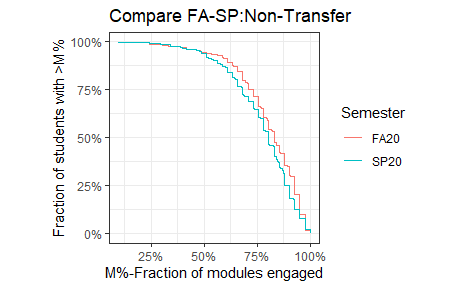}
            \label{fig: engage_betweenTransfer}
        }
    }    
    \caption{CCDs of fraction of engaged modules where a significant difference is detected within the same demographic group between FA20 and SP20 semesters: (a) Comparison for non-URM students. (b) Comparison for Non-transfer students.}
\end{figure}

Between semester comparisons (Table \ref{tab: between} column ``Engage") showed that non-URM students had significantly more engaged modules during FA20 semester than the SP20 semester (Figure \ref{fig: engage_betweenNURM}), whereas URM students remained the same. Non-transfer students are also observed to be more engaged with OLMs in FA20, but the difference is relatively small and did not result in observable gap between transfer and non-transfer students in either FA20 or SP20 semesters (Figure \ref{fig: engage_betweenTransfer}). 

\begin{table}[h!]
    \caption{Statistical test results comparing fraction of modules engaged between different demographics groups within the same semester.}
    \label{tab: engage_within}
    \centering
    \begin{tabular}{c c c c c c c}
        \toprule\toprule
         FA2020 & \multicolumn{2}{c}{FA2020} & \multicolumn{2}{c}{FA2020-A} & \multicolumn{2}{c}{SP2020-A} \\
         \arrayrulecolor{black!30}\midrule 
         Dimension & Statistic & p & Statistic & p & Statistic & p \\
         \arrayrulecolor{black}\midrule 
         Gender & 19276 & 0.15 & 4712 & 0.053 & 5860 & 0.09 \\
         \arrayrulecolor{black!30}\midrule 
         Frst.Gen & 14420 & 0.54 & 3880 & 0.62 & 3554 & 0.15 \\
         \midrule
         URM & 25076 & 0.93 & 6284 & 0.34 & \textbf{9789} & \textbf{0.004} \\
         \midrule
         Transfer & 20163 & 0.51 & 5399 & 0.54 & 3894 & 0.07 \\
         \arrayrulecolor{black}\bottomrule\bottomrule
    \end{tabular}
\end{table}


The linear model for FA20 semester found that female first generation students (14 students) had lower fraction of modules engaged compared to male non-first generation students($p < 0.05$). However, the statistical power of the overall model is limited (ANOVA test $p = 0.19$).

\begin{table}[h!]
    \caption{Summary of statistically significant differences observed in the current analysis. In the FA-SP Comparison column, (+) denotes a significant increase in a data indicator from SP20 to FA20 semester, whereas (-) denotes a decrease from SP20 to FA20.}
    \label{tab: sigDiff}
    \centering
    \begin{tabular}{l l l l}
        \toprule\toprule
         Data Indicator & FA20 & SP20 & FA-SP Comparison \\
         \arrayrulecolor{black}\midrule 
         On-Time Submissions & First Gen. $<$ Non-First Gen. & None & None \\
         \arrayrulecolor{black!30}\midrule 
         Early Submissions & First Gen. $<$ Non-First Gen & Transfer $<$ Non-Transfer & First Gen. (-) \\
         & URM $<$ Non-URM & & Non-First Gen (+) \\
         & & & Transfer (+) \\
         \midrule
         Engagement & None & URM $>$ Non-URM & Non-URM (+) \\
         \arrayrulecolor{black}\bottomrule\bottomrule
    \end{tabular}
\end{table}

\subsubsection{Comparing Summative Course Scores}
As shown in Table \ref{tab: grades_FA20}, in FA20, URM students had significantly lower quiz, exam, and overall course scores compared to non-URM students. Transfer students had lower quiz scores compared to non-transfer students. Similar differences regarding exam and course scores were also observed in the FA19 semester (Table \ref{tab: grades_FA19}). In FA20, first generation students had significantly lower overall course scores but not exam or quiz scores. Since this differences was not observed in FA19, the difference in FA20 is likely linked to the penalty from late submissions as well as significantly less extra credit for early submissions. In FA20, female students had significantly lower exam scores than male students, which was also not observed for FA19. Between semester comparisons of normalized total course scores across the two semesters did not identify any statistically significant difference. It should be noted that the exams in both semesters did not contain the same problems, and the weight towards total course grade is less in FA20 than in FA19, due to remote learning related adjustments to the course structure.

\begin{table}[h!]
    \caption{Statistical test results for comparison of summative course grades for FA20 semester between demographic groups.}
    \label{tab: grades_FA20}
    \centering
    \begin{tabular}{c c c c c c c}
        \toprule\toprule
         FA2020 & \multicolumn{2}{c}{Quiz} & \multicolumn{2}{c}{Exam} & \multicolumn{2}{c}{Course} \\
         \arrayrulecolor{black!30}\midrule 
         Dimension & Statistic & p & Statistic & p & Statistic & p \\
         \arrayrulecolor{black}\midrule 
         Gender & 5618 & 0.81 & \textbf{6620} & \textbf{0.02} & 5678 & 0.71 \\
         \arrayrulecolor{black!30}\midrule 
         Frst.Gen & 3372 & 0.16 & 3286 & 0.11 & \textbf{2998} & \textbf{0.02} \\
         \midrule
         URM & \textbf{5069} & \textbf{0.002} & \textbf{5521} & \textbf{0.03} & \textbf{4944} & \textbf{0.0008} \\
         \midrule
         Transfer & \textbf{6468} & \textbf{0.02} & 6132 & 0.11 & 6138 & 0.11\\
         \arrayrulecolor{black}\bottomrule\bottomrule
    \end{tabular}
\end{table}

\begin{table}[h!]
    \caption{Statistical test results for comparison of summative course grades for FA19 semester between demographic groups.}
    \label{tab: grades_FA19}
    \centering
    \begin{tabular}{c c c c c}
        \toprule\toprule
         FA2019 & \multicolumn{2}{c}{Exam} & \multicolumn{2}{c}{Course} \\
         \arrayrulecolor{black!30}\midrule 
         Dimension & Statistic & p & Statistic & p \\
         \arrayrulecolor{black}\midrule 
         Gender & 5865 & 0.44 & 5229 & 0.54 \\
         \arrayrulecolor{black!30}\midrule 
         Frst.Gen & 5578 & 0.66 & 5857 & 0.88 \\
         \midrule
         URM & \textbf{1776} & \textbf{0.002} & \textbf{2071} & \textbf{0.009} \\
         \midrule
         Transfer & \textbf{9168} & \textbf{0.0013} & \textbf{8724} & \textbf{0.02} \\
         \arrayrulecolor{black}\bottomrule\bottomrule
    \end{tabular}
\end{table}

\section{Discussion}\label{sec: discussion}
Our analysis suggests that the overall impact of the pandemic remote learning on students' ability to keep up with the pace of instruction is relatively small. Both on-time and early submission frequencies are identical in F2F and remote learning settings and the fraction of engaged modules increased slightly during remote learning.

On the other hand, the impact of remote learning is heterogeneous across demographic divisions which is in agreement with previous studies \cite{Aucejo2020, Gillis2020}. Statistically significant differences both between the two demographic groups in the same semester, and within the same group between different semesters, are summarised in Table \ref{tab: sigDiff}. We found that first generation students were significantly less likely to submit the OLMs either on-time or early compared to both non-first generation students in remote learning and first generation students in an F2F setting. One possible explanation is that first generation students tend to rely much more on physical infrastructure and resources provided by the institutions for learning, and were faced with more distractions, challenges, and family responsibilities while studying from home. Meanwhile, some non-first generation students may have access to better learning environments at home, resulting in an increase in frequency of early OLM submission.

On the contrary, we observed that transfer students' frequency of early submission increased during the remote instruction period to a similar level as non-transfer students, whereas during F2F instruction transfer students had significantly lower fraction of early submission. Since the majority of transfer students came from small two-year local state colleges, it is possible that many faced challenges adjusting to life at a large public four-year university. Remote learning might have served to reduce the level of such "cultural shock" and facilitated the transfer students in planning their learning schedules.

Regarding engagement with the OLMs, we found that in SP20, URM students were more engaged when compared to non-URM students. However, this difference was not observed in FA20 remote instruction due to an increase in the level of engagement among non-URM students. More future research is needed to reveal the possible causes of this observation and implications for instruction. On the other hand, while first generation students tend to submit OLMs later than their peers, they are equally engaged with learning from those modules. This is in agreement with the observation that first generation students had similar quiz and exam scores as non-first generation students, but lower overall course grades which is affected by late submission penalties and early submissions extra credits. 

Moreover, it is alarming that while URM and transfer students are engaging with and completing OLMs at the same or higher rate compared to their peers, their exam scores and final course outcomes are significantly worse in both F2F and remote instruction. In an earlier study, Salehi et al. \cite{Salehi2019} found that demographic gaps in exam scores could be completely explained by difference in incoming preparation. In future studies, we will investigate whether the gaps in the current physics course are also related to incoming preparation gaps and whether remote learning increases or reduces those gaps.

Another earlier study \cite{Aucejo2020} suggests that the impacts of pandemic remote learning on different student demographics are mediated by differences in socioeconomic status. While we do not have access to data on individual student's family income status, we did compare the fraction of PELL-eligible students in each demographic group. In both semesters,  first generation students have significantly higher fractions of PELL-eligibility compared to non-first generation students, which might help explain their decline in on-time and early submission frequency. However, transfer students also had significantly higher fraction of PELL-eligibility than non-transfer students in both semesters, yet their early submission frequency increased during remote learning. Finally, while non-URM students had a higher fraction of engaged modules in FA20 compared to SP20, the fraction of PELL-Eligible non-URM students is actually significantly higher in FA20 compared to SP20. From those results it seems that socioeconomic status is likely not a major mediator (or at the very least not the only mediator) of the impact of remote learning for all demographic groups.

Finally, we address a technical issue regarding statistical analysis. For each of the three data indicators in FA20 and SP20, four statistical tests were conducted to compare between four different pairs of demographic groups. Strictly speaking, conducting multiple tests increases the likelihood of finding a significant difference by chance (Type I error), and to compensate for this increase the p-values need to be adjusted \cite{Jafari}. While p-value adjustments are more commonly conducted for studies involving hundreds of tests, we nevertheless re-examined our results presented in Tables \ref{tab: onTime_within}, \ref{tab: early_within} and \ref{tab: engage_within} after conducting p-value adjustment using the Banjamini-Hochberg method. \cite{Benjamini1995} Most significant differences remain significant at $\alpha = 0.05$ level, except the differences in fraction of on-time submission between first generation and non-first generation students for both FA20 all modules and first 41 modules (\ref{tab: onTime_within}). This change has little impact on the main conclusion of the paper since a larger difference of the same nature is observed for early submission. In addition, the differences between URM and non-URM students in fraction early submission for the first 41 modules in FA20 were also no longer significant. However, the original difference was small in the first place, and was not observed for the entire semester. We did not conduct p-value adjustment for the between semester comparison results in Table \ref{tab: between}, because we were looking for significant differences withing one or two specific groups (such as First Generation students), rather than searching for any significant differences. Conducting multiple tests will not increase the Type I error probability for a particular test.

\section{Conclusion and Future Directions}

The key takeaway from the current analysis is that the same change in instructional conditions could have very different impacts on different groups of students, sometimes even in opposite directions. Therefore as we emerge from the COVID pandemic, the discussion on the future of STEM education should no longer be centered on whether one mode of instruction is "better" than another, but rather on how to provide more flexible options to meet the increasingly diverse needs of the student population \cite{U.SDepartmentofEducation2017}. Moreover, decisions on instructional design should be based increasingly on detailed data analysis and careful student modeling that takes into account the diversity of the students' background, rather than modeling the "average student".

The current analysis serves as only a starting point for investigation into the heterogeneous impact of online and remote instruction on different student bodies. A number of major improvements can be made in future follow up studies. First, the data indicators chosen in the current study only serve as coarse approximations of the complicated student learning behavior in an online environment. Future studies should model students' behavior much more carefully by employing advanced analysis methods such as process mining \cite{Maldonado-Mahauad2018, Saint2021} and grounding the analysis in established theoretical frameworks such as self-regulated learning \cite{Zimmerman2000, Zimmerman2013}. Secondly, while the sample size ({\raise.17ex\hbox{$\scriptstyle\sim$}}250) of the current study is sufficient to measure the impact on minority groups based on single demographic variables, larger sample size in future studies are needed to obtaining statistically meaningful results for the intersectional groups (e.g., female-URM or transfer-first generation students), by conducting more detailed linear modeling. Furthermore, the same analysis should be repeated on data from multiple future semesters to examine the possibility that the observed differences arise from fluctuation in student population unrelated to pandemic remote learning. Finally, several earlier studies \cite{Gillis2020, Orlov2021} pointed out that differences in instructional strategies and implementation of instruction can have significant impact on students' perception of remote learning. While the current study investigated students' behavior in one class using one form on online homework, future studies should compare the impacts different types of instructional designs and other choices of educational technologies on students' learning behavior. 


\acknowledgments{This work is supported by NSF Award No. DUE-1845436. We would like to thank the UCF Learning Systems and Technologies team led by Dr. Francisca Yonekura for developing the Obojobo platform; Dr. Mengyu Xu for valuable input on the linear models, and Zachary Felker for analysis of students' submission schedule.}

\section*{References}
\bibliography{CovidRelated.bib}

\begin{thebibliography}{33}%
\makeatletter
\providecommand \@ifxundefined [1]{%
 \@ifx{#1\undefined}
}%
\providecommand \@ifnum [1]{%
 \ifnum #1\expandafter \@firstoftwo
 \else \expandafter \@secondoftwo
 \fi
}%
\providecommand \@ifx [1]{%
 \ifx #1\expandafter \@firstoftwo
 \else \expandafter \@secondoftwo
 \fi
}%
\providecommand \natexlab [1]{#1}%
\providecommand \enquote  [1]{``#1''}%
\providecommand \bibnamefont  [1]{#1}%
\providecommand \bibfnamefont [1]{#1}%
\providecommand \citenamefont [1]{#1}%
\providecommand \href@noop [0]{\@secondoftwo}%
\providecommand \href [0]{\begingroup \@sanitize@url \@href}%
\providecommand \@href[1]{\@@startlink{#1}\@@href}%
\providecommand \@@href[1]{\endgroup#1\@@endlink}%
\providecommand \@sanitize@url [0]{\catcode `\\12\catcode `\$12\catcode
  `\&12\catcode `\#12\catcode `\^12\catcode `\_12\catcode `\%12\relax}%
\providecommand \@@startlink[1]{}%
\providecommand \@@endlink[0]{}%
\providecommand \url  [0]{\begingroup\@sanitize@url \@url }%
\providecommand \@url [1]{\endgroup\@href {#1}{\urlprefix }}%
\providecommand \urlprefix  [0]{URL }%
\providecommand \Eprint [0]{\href }%
\providecommand \doibase [0]{https://doi.org/}%
\providecommand \selectlanguage [0]{\@gobble}%
\providecommand \bibinfo  [0]{\@secondoftwo}%
\providecommand \bibfield  [0]{\@secondoftwo}%
\providecommand \translation [1]{[#1]}%
\providecommand \BibitemOpen [0]{}%
\providecommand \bibitemStop [0]{}%
\providecommand \bibitemNoStop [0]{.\EOS\space}%
\providecommand \EOS [0]{\spacefactor3000\relax}%
\providecommand \BibitemShut  [1]{\csname bibitem#1\endcsname}%
\let\auto@bib@innerbib\@empty
\bibitem [{\citenamefont {Biwer}\ \emph {et~al.}(2021)\citenamefont {Biwer},
  \citenamefont {Wiradhany}, \citenamefont {oude Egbrink}, \citenamefont
  {Hospers}, \citenamefont {Wasenitz}, \citenamefont {Jansen},\ and\
  \citenamefont {de~Bruin}}]{Biwer2021}%
  \BibitemOpen
  \bibfield  {author} {\bibinfo {author} {\bibfnamefont {F.}~\bibnamefont
  {Biwer}}, \bibinfo {author} {\bibfnamefont {W.}~\bibnamefont {Wiradhany}},
  \bibinfo {author} {\bibfnamefont {M.}~\bibnamefont {oude Egbrink}}, \bibinfo
  {author} {\bibfnamefont {H.}~\bibnamefont {Hospers}}, \bibinfo {author}
  {\bibfnamefont {S.}~\bibnamefont {Wasenitz}}, \bibinfo {author}
  {\bibfnamefont {W.}~\bibnamefont {Jansen}},\ and\ \bibinfo {author}
  {\bibfnamefont {A.}~\bibnamefont {de~Bruin}},\ }\bibfield  {title} {\enquote
  {\bibinfo {title} {Changes and adaptations: How university students
  self-regulate their online learning during the covid-19 pandemic},}\ }\href
  {https://doi.org/10.3389/fpsyg.2021.642593} {\bibfield  {journal} {\bibinfo
  {journal} {Frontiers in Psychology}\ }\textbf {\bibinfo {volume} {12}}
  (\bibinfo {year} {2021}),\ 10.3389/fpsyg.2021.642593}\BibitemShut {NoStop}%
\bibitem [{\citenamefont {Orlov}\ \emph {et~al.}(2021)\citenamefont {Orlov},
  \citenamefont {McKee}, \citenamefont {Berry}, \citenamefont {Boyle},
  \citenamefont {DiCiccio}, \citenamefont {Ransom}, \citenamefont
  {Rees-Jones},\ and\ \citenamefont {Stoye}}]{Orlov2021}%
  \BibitemOpen
  \bibfield  {author} {\bibinfo {author} {\bibfnamefont {G.}~\bibnamefont
  {Orlov}}, \bibinfo {author} {\bibfnamefont {D.}~\bibnamefont {McKee}},
  \bibinfo {author} {\bibfnamefont {J.}~\bibnamefont {Berry}}, \bibinfo
  {author} {\bibfnamefont {A.}~\bibnamefont {Boyle}}, \bibinfo {author}
  {\bibfnamefont {T.}~\bibnamefont {DiCiccio}}, \bibinfo {author}
  {\bibfnamefont {T.}~\bibnamefont {Ransom}}, \bibinfo {author} {\bibfnamefont
  {A.}~\bibnamefont {Rees-Jones}},\ and\ \bibinfo {author} {\bibfnamefont
  {J.}~\bibnamefont {Stoye}},\ }\bibfield  {title} {\enquote {\bibinfo {title}
  {{Learning during the COVID-19 pandemic: It is not who you teach, but how you
  teach}},}\ }\href {https://doi.org/10.1016/j.econlet.2021.109812} {\bibfield
  {journal} {\bibinfo  {journal} {Economics Letters}\ }\textbf {\bibinfo
  {volume} {202}},\ \bibinfo {pages} {109812} (\bibinfo {year}
  {2021})}\BibitemShut {NoStop}%
\bibitem [{\citenamefont {Daniels}, \citenamefont {Goegan},\ and\ \citenamefont
  {Parker}(2021)}]{Daniels2021}%
  \BibitemOpen
  \bibfield  {author} {\bibinfo {author} {\bibfnamefont {L.~M.}\ \bibnamefont
  {Daniels}}, \bibinfo {author} {\bibfnamefont {L.~D.}\ \bibnamefont
  {Goegan}},\ and\ \bibinfo {author} {\bibfnamefont {P.~C.}\ \bibnamefont
  {Parker}},\ }\bibfield  {title} {\enquote {\bibinfo {title} {{The impact of
  COVID-19 triggered changes to instruction and assessment on university
  students' self-reported motivation, engagement and perceptions}},}\ }\href
  {https://doi.org/10.1007/s11218-021-09612-3} {\bibfield  {journal} {\bibinfo
  {journal} {Social Psychology of Education}\ }\textbf {\bibinfo {volume}
  {24}},\ \bibinfo {pages} {299--318} (\bibinfo {year} {2021})}\BibitemShut
  {NoStop}%
\bibitem [{\citenamefont {Elmer}, \citenamefont {Mepham},\ and\ \citenamefont
  {Stadtfeld}(2020)}]{Elmer2020}%
  \BibitemOpen
  \bibfield  {author} {\bibinfo {author} {\bibfnamefont {T.}~\bibnamefont
  {Elmer}}, \bibinfo {author} {\bibfnamefont {K.}~\bibnamefont {Mepham}},\ and\
  \bibinfo {author} {\bibfnamefont {C.}~\bibnamefont {Stadtfeld}},\ }\bibfield
  {title} {\enquote {\bibinfo {title} {{Students under lockdown: Comparisons of
  students' social networks and mental health before and during the COVID-19
  crisis in Switzerland}},}\ }\href
  {https://doi.org/10.1371/journal.pone.0236337} {\bibfield  {journal}
  {\bibinfo  {journal} {PLoS ONE}\ }\textbf {\bibinfo {volume} {15}},\ \bibinfo
  {pages} {1--22} (\bibinfo {year} {2020})}\BibitemShut {NoStop}%
\bibitem [{\citenamefont {Aucejo}\ \emph {et~al.}(2020)\citenamefont {Aucejo},
  \citenamefont {French}, \citenamefont {{Ugalde Araya}},\ and\ \citenamefont
  {Zafar}}]{Aucejo2020}%
  \BibitemOpen
  \bibfield  {author} {\bibinfo {author} {\bibfnamefont {E.~M.}\ \bibnamefont
  {Aucejo}}, \bibinfo {author} {\bibfnamefont {J.}~\bibnamefont {French}},
  \bibinfo {author} {\bibfnamefont {M.~P.}\ \bibnamefont {{Ugalde Araya}}},\
  and\ \bibinfo {author} {\bibfnamefont {B.}~\bibnamefont {Zafar}},\ }\bibfield
   {title} {\enquote {\bibinfo {title} {{The impact of COVID-19 on student
  experiences and expectations: Evidence from a survey}},}\ }\href
  {https://doi.org/10.1016/j.jpubeco.2020.104271} {\bibfield  {journal}
  {\bibinfo  {journal} {Journal of Public Economics}\ }\textbf {\bibinfo
  {volume} {191}},\ \bibinfo {pages} {104271} (\bibinfo {year}
  {2020})}\BibitemShut {NoStop}%
\bibitem [{\citenamefont {El~Said}(2021)}]{ElSaid2021}%
  \BibitemOpen
  \bibfield  {author} {\bibinfo {author} {\bibfnamefont {G.~R.}\ \bibnamefont
  {El~Said}},\ }\bibfield  {title} {\enquote {\bibinfo {title} {{How Did the
  COVID-19 Pandemic Affect Higher Education Learning Experience? An Empirical
  Investigation of Learners' Academic Performance at a University in a
  Developing Country}},}\ }\href {https://doi.org/10.1155/2021/6649524}
  {\bibfield  {journal} {\bibinfo  {journal} {Advances in Human-Computer
  Interaction}\ }\textbf {\bibinfo {volume} {2021}} (\bibinfo {year} {2021}),\
  10.1155/2021/6649524}\BibitemShut {NoStop}%
\bibitem [{\citenamefont {Gonzalez}\ \emph {et~al.}(2020)\citenamefont
  {Gonzalez}, \citenamefont {{De la Rubia}}, \citenamefont {Hincz},
  \citenamefont {Comas-Lopez}, \citenamefont {Subirats}, \citenamefont {Fort},\
  and\ \citenamefont {Sacha}}]{Gonzalez2020}%
  \BibitemOpen
  \bibfield  {author} {\bibinfo {author} {\bibfnamefont {T.}~\bibnamefont
  {Gonzalez}}, \bibinfo {author} {\bibfnamefont {M.~A.}\ \bibnamefont {{De la
  Rubia}}}, \bibinfo {author} {\bibfnamefont {K.~P.}\ \bibnamefont {Hincz}},
  \bibinfo {author} {\bibfnamefont {M.}~\bibnamefont {Comas-Lopez}}, \bibinfo
  {author} {\bibfnamefont {L.}~\bibnamefont {Subirats}}, \bibinfo {author}
  {\bibfnamefont {S.}~\bibnamefont {Fort}},\ and\ \bibinfo {author}
  {\bibfnamefont {G.~M.}\ \bibnamefont {Sacha}},\ }\bibfield  {title} {\enquote
  {\bibinfo {title} {{Influence of COVID-19 confinement on students'
  performance in higher education}},}\ }\href
  {https://doi.org/10.1371/journal.pone.0239490} {\bibfield  {journal}
  {\bibinfo  {journal} {PLoS ONE}\ }\textbf {\bibinfo {volume} {15}},\ \bibinfo
  {pages} {1--25} (\bibinfo {year} {2020})},\ \Eprint
  {https://arxiv.org/abs/2004.09545} {arXiv:2004.09545} \BibitemShut {NoStop}%
\bibitem [{\citenamefont {Wilcox}\ and\ \citenamefont
  {Vignal}(2020)}]{Wilcox2020}%
  \BibitemOpen
  \bibfield  {author} {\bibinfo {author} {\bibfnamefont {B.~R.}\ \bibnamefont
  {Wilcox}}\ and\ \bibinfo {author} {\bibfnamefont {M.}~\bibnamefont
  {Vignal}},\ }\bibfield  {title} {\enquote {\bibinfo {title} {{Understanding
  the student experience with emergency remote teaching}},}\ }in\ \href
  {https://doi.org/10.1119/perc.2020.pr.wilcox} {\emph {\bibinfo {booktitle}
  {2020 Physics Education Research Conference Proceedings}}}\ (\bibinfo
  {publisher} {American Association of Physics Teachers (AAPT)},\ \bibinfo
  {address} {Virtual Conference},\ \bibinfo {year} {2020})\ pp.\ \bibinfo
  {pages} {581--586}\BibitemShut {NoStop}%
\bibitem [{\citenamefont {Fox}\ \emph {et~al.}(2021)\citenamefont {Fox},
  \citenamefont {Hoehn}, \citenamefont {Werth},\ and\ \citenamefont
  {Lewandowski}}]{Fox2021}%
  \BibitemOpen
  \bibfield  {author} {\bibinfo {author} {\bibfnamefont {M.~F.~J.}\
  \bibnamefont {Fox}}, \bibinfo {author} {\bibfnamefont {J.~R.}\ \bibnamefont
  {Hoehn}}, \bibinfo {author} {\bibfnamefont {A.}~\bibnamefont {Werth}},\ and\
  \bibinfo {author} {\bibfnamefont {H.~J.}\ \bibnamefont {Lewandowski}},\
  }\bibfield  {title} {\enquote {\bibinfo {title} {{Lab instruction during the
  COVID-19 pandemic: Effects on student views about experimental physics in
  comparison with previous years}},}\ }\href
  {https://doi.org/10.1103/physrevphyseducres.17.010148} {\bibfield  {journal}
  {\bibinfo  {journal} {Physical Review Physics Education Research}\ }\textbf
  {\bibinfo {volume} {17}},\ \bibinfo {pages} {10148} (\bibinfo {year}
  {2021})}\BibitemShut {NoStop}%
\bibitem [{\citenamefont {Klein}\ \emph {et~al.}(2021)\citenamefont {Klein},
  \citenamefont {Ivanjek}, \citenamefont {Dahlkemper}, \citenamefont
  {Jeli{\v{c}}i{\'{c}}}, \citenamefont {Geyer}, \citenamefont
  {K{\"{u}}chemann},\ and\ \citenamefont {Susac}}]{Klein2021}%
  \BibitemOpen
  \bibfield  {author} {\bibinfo {author} {\bibfnamefont {P.}~\bibnamefont
  {Klein}}, \bibinfo {author} {\bibfnamefont {L.}~\bibnamefont {Ivanjek}},
  \bibinfo {author} {\bibfnamefont {M.~N.}\ \bibnamefont {Dahlkemper}},
  \bibinfo {author} {\bibfnamefont {K.}~\bibnamefont {Jeli{\v{c}}i{\'{c}}}},
  \bibinfo {author} {\bibfnamefont {M.~A.}\ \bibnamefont {Geyer}}, \bibinfo
  {author} {\bibfnamefont {S.}~\bibnamefont {K{\"{u}}chemann}},\ and\ \bibinfo
  {author} {\bibfnamefont {A.}~\bibnamefont {Susac}},\ }\bibfield  {title}
  {\enquote {\bibinfo {title} {{Studying physics during the COVID-19 pandemic:
  Student assessments of learning achievement, perceived effectiveness of
  online recitations, and online laboratories}},}\ }\href
  {https://doi.org/10.1103/PhysRevPhysEducRes.17.010117} {\bibfield  {journal}
  {\bibinfo  {journal} {Physical Review Physics Education Research}\ }\textbf
  {\bibinfo {volume} {17}},\ \bibinfo {pages} {1--11} (\bibinfo {year}
  {2021})},\ \Eprint {https://arxiv.org/abs/2010.05622} {arXiv:2010.05622}
  \BibitemShut {NoStop}%
\bibitem [{\citenamefont {Gillis}\ and\ \citenamefont
  {Krull}(2020)}]{Gillis2020}%
  \BibitemOpen
  \bibfield  {author} {\bibinfo {author} {\bibfnamefont {A.}~\bibnamefont
  {Gillis}}\ and\ \bibinfo {author} {\bibfnamefont {L.~M.}\ \bibnamefont
  {Krull}},\ }\bibfield  {title} {\enquote {\bibinfo {title} {{COVID-19 Remote
  Learning Transition in Spring 2020: Class Structures, Student Perceptions,
  and Inequality in College Courses}},}\ }\href
  {https://doi.org/10.1177/0092055X20954263} {\bibfield  {journal} {\bibinfo
  {journal} {Teaching Sociology}\ }\textbf {\bibinfo {volume} {48}},\ \bibinfo
  {pages} {283--299} (\bibinfo {year} {2020})}\BibitemShut {NoStop}%
\bibitem [{\citenamefont {Naujoks}\ \emph {et~al.}(2021)\citenamefont
  {Naujoks}, \citenamefont {Bedenlier}, \citenamefont {Gl{\"{a}}ser-Zikuda},
  \citenamefont {Kammerl}, \citenamefont {Kopp}, \citenamefont {Ziegler},\ and\
  \citenamefont {H{\"{a}}ndel}}]{Naujoks2021}%
  \BibitemOpen
  \bibfield  {author} {\bibinfo {author} {\bibfnamefont {N.}~\bibnamefont
  {Naujoks}}, \bibinfo {author} {\bibfnamefont {S.}~\bibnamefont {Bedenlier}},
  \bibinfo {author} {\bibfnamefont {M.}~\bibnamefont {Gl{\"{a}}ser-Zikuda}},
  \bibinfo {author} {\bibfnamefont {R.}~\bibnamefont {Kammerl}}, \bibinfo
  {author} {\bibfnamefont {B.}~\bibnamefont {Kopp}}, \bibinfo {author}
  {\bibfnamefont {A.}~\bibnamefont {Ziegler}},\ and\ \bibinfo {author}
  {\bibfnamefont {M.}~\bibnamefont {H{\"{a}}ndel}},\ }\bibfield  {title}
  {\enquote {\bibinfo {title} {{Self-Regulated Resource Management in Emergency
  Remote Higher Education: Status Quo and Predictors}},}\ }\href
  {https://doi.org/10.3389/fpsyg.2021.672741} {\bibfield  {journal} {\bibinfo
  {journal} {Frontiers in Psychology}\ }\textbf {\bibinfo {volume} {12}}
  (\bibinfo {year} {2021}),\ 10.3389/fpsyg.2021.672741}\BibitemShut {NoStop}%
\bibitem [{\citenamefont {Chhetri}(2020)}]{Chhetri2020}%
  \BibitemOpen
  \bibfield  {author} {\bibinfo {author} {\bibfnamefont {C.}~\bibnamefont
  {Chhetri}},\ }\bibfield  {title} {\enquote {\bibinfo {title} {{"I Lost Track
  of Things": Student Experiences of Remote Learning in the Covid-19
  Pandemic}},}\ }\href {https://doi.org/10.1145/3368308.3415413} {\bibfield
  {journal} {\bibinfo  {journal} {SIGITE 2020 - Proceedings of the 21st Annual
  Conference on Information Technology Education}\ ,\ \bibinfo {pages}
  {314--319}} (\bibinfo {year} {2020})}\BibitemShut {NoStop}%
\bibitem [{\citenamefont {Zhang}, \citenamefont {Taub},\ and\ \citenamefont
  {Chen}(2021)}]{Zhang2021}%
  \BibitemOpen
  \bibfield  {author} {\bibinfo {author} {\bibfnamefont {T.}~\bibnamefont
  {Zhang}}, \bibinfo {author} {\bibfnamefont {M.}~\bibnamefont {Taub}},\ and\
  \bibinfo {author} {\bibfnamefont {Z.}~\bibnamefont {Chen}},\ }\enquote
  {\bibinfo {title} {Measuring the impact of covid-19 induced campus closure on
  student self-regulated learning in physics online learning modules},}\ in\
  \href {https://doi.org/10.1145/3448139.3448150} {\emph {\bibinfo {booktitle}
  {LAK21: 11th International Learning Analytics and Knowledge Conference}}}\
  (\bibinfo  {publisher} {Association for Computing Machinery},\ \bibinfo
  {address} {New York, NY, USA},\ \bibinfo {year} {2021})\ p.\ \bibinfo {pages}
  {110–120}\BibitemShut {NoStop}%
\bibitem [{\citenamefont {Chen}\ \emph {et~al.}(2020)\citenamefont {Chen},
  \citenamefont {Xu}, \citenamefont {Garrido},\ and\ \citenamefont
  {Guthrie}}]{Chen2020}%
  \BibitemOpen
  \bibfield  {author} {\bibinfo {author} {\bibfnamefont {Z.}~\bibnamefont
  {Chen}}, \bibinfo {author} {\bibfnamefont {M.}~\bibnamefont {Xu}}, \bibinfo
  {author} {\bibfnamefont {G.}~\bibnamefont {Garrido}},\ and\ \bibinfo {author}
  {\bibfnamefont {M.~W.}\ \bibnamefont {Guthrie}},\ }\bibfield  {title}
  {\enquote {\bibinfo {title} {{Relationship between students' online learning
  behavior and course performance: What contextual information matters?}}}\
  }\href {https://doi.org/10.1103/PhysRevPhysEducRes.16.010138} {\bibfield
  {journal} {\bibinfo  {journal} {Physical Review Physics Education Research}\
  }\textbf {\bibinfo {volume} {16}},\ \bibinfo {pages} {010138} (\bibinfo
  {year} {2020})}\BibitemShut {NoStop}%
\bibitem [{\citenamefont {Chen}(2021)}]{Chen2021}%
  \BibitemOpen
  \bibfield  {author} {\bibinfo {author} {\bibfnamefont {Z.}~\bibnamefont
  {Chen}},\ }\href {https://www.aaas-iuse.org/resource/course-design/}
  {\enquote {\bibinfo {title} {{How the Abrupt Shift Online Led to a Permanent
  Shift in Course Design}},}\ } (\bibinfo {year} {2021})\BibitemShut {NoStop}%
\bibitem [{\citenamefont {Chen}, \citenamefont {Lee},\ and\ \citenamefont
  {Garrido}(2018)}]{Chen2018}%
  \BibitemOpen
  \bibfield  {author} {\bibinfo {author} {\bibfnamefont {Z.}~\bibnamefont
  {Chen}}, \bibinfo {author} {\bibfnamefont {S.}~\bibnamefont {Lee}},\ and\
  \bibinfo {author} {\bibfnamefont {G.}~\bibnamefont {Garrido}},\ }\bibfield
  {title} {\enquote {\bibinfo {title} {{Re-designing the Structure of Online
  Courses to Empower Educational Data Mining}},}\ }in\ \href@noop {} {\emph
  {\bibinfo {booktitle} {Proceedings of 11th International Educational Data
  Mining Conference}}},\ \bibinfo {editor} {edited by\ \bibinfo {editor}
  {\bibfnamefont {K.}~\bibnamefont {{Elizabeth Boyer}}}\ and\ \bibinfo {editor}
  {\bibfnamefont {M.}~\bibnamefont {Yudelson}}}\ (\bibinfo {address} {Buffalo,
  NY},\ \bibinfo {year} {2018})\ pp.\ \bibinfo {pages} {390--396}\BibitemShut
  {NoStop}%
\bibitem [{\citenamefont {Chen}\ \emph {et~al.}(2018)\citenamefont {Chen},
  \citenamefont {Garrido}, \citenamefont {Berry}, \citenamefont {Turgeon},\
  and\ \citenamefont {Yonekura}}]{Chen2018a}%
  \BibitemOpen
  \bibfield  {author} {\bibinfo {author} {\bibfnamefont {Z.}~\bibnamefont
  {Chen}}, \bibinfo {author} {\bibfnamefont {G.}~\bibnamefont {Garrido}},
  \bibinfo {author} {\bibfnamefont {Z.}~\bibnamefont {Berry}}, \bibinfo
  {author} {\bibfnamefont {I.}~\bibnamefont {Turgeon}},\ and\ \bibinfo {author}
  {\bibfnamefont {F.}~\bibnamefont {Yonekura}},\ }\bibfield  {title} {\enquote
  {\bibinfo {title} {{Designing online learning modules to conduct pre- and
  post-testing at high frequency}},}\ }in\ \href
  {https://doi.org/10.1119/perc.2017.pr.016} {\emph {\bibinfo {booktitle} {2017
  Physics Education Research Conference Proceedings}}}\ (\bibinfo  {publisher}
  {American Association of Physics Teachers},\ \bibinfo {address} {Cincinnati,
  OH},\ \bibinfo {year} {2018})\ pp.\ \bibinfo {pages} {84--87}\BibitemShut
  {NoStop}%
\bibitem [{\citenamefont {Felker}\ and\ \citenamefont
  {Chen}(2020)}]{Felker2020}%
  \BibitemOpen
  \bibfield  {author} {\bibinfo {author} {\bibfnamefont {Z.}~\bibnamefont
  {Felker}}\ and\ \bibinfo {author} {\bibfnamefont {Z.}~\bibnamefont {Chen}},\
  }\bibfield  {title} {\enquote {\bibinfo {title} {{The impact of extra credit
  incentives on students' work habits when completing online homework
  assignments}},}\ }in\ \href {https://doi.org/10.1119/perc.2020.pr.felker}
  {\emph {\bibinfo {booktitle} {2020 Physics Education Research Conference
  Proceedings}}}\ (\bibinfo  {publisher} {American Association of Physics
  Teachers (AAPT)},\ \bibinfo {address} {Virtual Conference},\ \bibinfo {year}
  {2020})\ pp.\ \bibinfo {pages} {143--148}\BibitemShut {NoStop}%
\bibitem [{\citenamefont {Guthrie}, \citenamefont {Zhang},\ and\ \citenamefont
  {Chen}(2020)}]{Guthrie2020}%
  \BibitemOpen
  \bibfield  {author} {\bibinfo {author} {\bibfnamefont {M.~W.}\ \bibnamefont
  {Guthrie}}, \bibinfo {author} {\bibfnamefont {T.}~\bibnamefont {Zhang}},\
  and\ \bibinfo {author} {\bibfnamefont {Z.}~\bibnamefont {Chen}},\ }\bibfield
  {title} {\enquote {\bibinfo {title} {{A tale of two guessing strategies:
  interpreting the time students spend solving problems through online log
  data}},}\ }in\ \href {https://doi.org/10.1119/perc.2020.pr.guthrie} {\emph
  {\bibinfo {booktitle} {Physics Education Research Conference Proceedings}}}\
  (\bibinfo  {publisher} {American Association of Physics Teachers (AAPT)},\
  \bibinfo {address} {Virtual Conference},\ \bibinfo {year} {2020})\ pp.\
  \bibinfo {pages} {185--190}\BibitemShut {NoStop}%
\bibitem [{\citenamefont {Seaton}\ \emph
  {et~al.}(2014{\natexlab{a}})\citenamefont {Seaton}, \citenamefont {Bergner},
  \citenamefont {Kortemeyer}, \citenamefont {Rayyan}, \citenamefont {Chuang},\
  and\ \citenamefont {Pritchard}}]{Seaton2014}%
  \BibitemOpen
  \bibfield  {author} {\bibinfo {author} {\bibfnamefont {D.~T.}\ \bibnamefont
  {Seaton}}, \bibinfo {author} {\bibfnamefont {Y.}~\bibnamefont {Bergner}},
  \bibinfo {author} {\bibfnamefont {G.}~\bibnamefont {Kortemeyer}}, \bibinfo
  {author} {\bibfnamefont {S.}~\bibnamefont {Rayyan}}, \bibinfo {author}
  {\bibfnamefont {I.}~\bibnamefont {Chuang}},\ and\ \bibinfo {author}
  {\bibfnamefont {D.~E.}\ \bibnamefont {Pritchard}},\ }\bibfield  {title}
  {\enquote {\bibinfo {title} {{The Impact of Course Structure on eText Use in
  Large-Lecture Introductory-Physics Courses}},}\ }\href
  {https://doi.org/10.1119/perc.2013.pr.071} {\bibfield  {journal} {\bibinfo
  {journal} {2013 Physics Education Research Conference Proceedings}\ ,\
  \bibinfo {pages} {333--336}} (\bibinfo {year}
  {2014}{\natexlab{a}})}\BibitemShut {NoStop}%
\bibitem [{\citenamefont {Seaton}\ \emph
  {et~al.}(2014{\natexlab{b}})\citenamefont {Seaton}, \citenamefont {Bergner},
  \citenamefont {Chuang}, \citenamefont {Mitros},\ and\ \citenamefont
  {Pritchard}}]{Seaton2014_2}%
  \BibitemOpen
  \bibfield  {author} {\bibinfo {author} {\bibfnamefont {D.~T.}\ \bibnamefont
  {Seaton}}, \bibinfo {author} {\bibfnamefont {Y.}~\bibnamefont {Bergner}},
  \bibinfo {author} {\bibfnamefont {I.}~\bibnamefont {Chuang}}, \bibinfo
  {author} {\bibfnamefont {P.}~\bibnamefont {Mitros}},\ and\ \bibinfo {author}
  {\bibfnamefont {D.~E.}\ \bibnamefont {Pritchard}},\ }\bibfield  {title}
  {\enquote {\bibinfo {title} {{Who does what in a massive open online
  course?}}}\ }\href {https://doi.org/10.1145/2500876} {\bibfield  {journal}
  {\bibinfo  {journal} {Communications of the ACM}\ }\textbf {\bibinfo {volume}
  {57}},\ \bibinfo {pages} {58--65} (\bibinfo {year}
  {2014}{\natexlab{b}})}\BibitemShut {NoStop}%
\bibitem [{\citenamefont {{Center for Distributed
  Learning}}()}]{CenterforDistributedLearning}%
  \BibitemOpen
  \bibfield  {author} {\bibinfo {author} {\bibnamefont {{Center for Distributed
  Learning}}},\ }\href {https://next.obojobo.ucf.edu/} {\enquote {\bibinfo
  {title} {{Obojobo}},}\ }\BibitemShut {NoStop}%
\bibitem [{\citenamefont {{R Core Team}}(2021)}]{R}%
  \BibitemOpen
  \bibfield  {author} {\bibinfo {author} {\bibnamefont {{R Core Team}}},\
  }\href {https://www.R-project.org/} {\emph {\bibinfo {title} {R: A Language
  and Environment for Statistical Computing}}},\ \bibinfo {organization} {R
  Foundation for Statistical Computing},\ \bibinfo {address} {Vienna, Austria}
  (\bibinfo {year} {2021})\BibitemShut {NoStop}%
\bibitem [{\citenamefont {Wickham}\ \emph {et~al.}(2019)\citenamefont
  {Wickham}, \citenamefont {Averick}, \citenamefont {Bryan}, \citenamefont
  {Chang}, \citenamefont {McGowan}, \citenamefont {Fran{\c{c}}ois},
  \citenamefont {Grolemund}, \citenamefont {Hayes}, \citenamefont {Henry},
  \citenamefont {Hester}, \citenamefont {Kuhn}, \citenamefont {Pedersen},
  \citenamefont {Miller}, \citenamefont {Bache}, \citenamefont {M{\"{u}}ller},
  \citenamefont {Ooms}, \citenamefont {Robinson}, \citenamefont {Seidel},
  \citenamefont {Spinu}, \citenamefont {Takahashi}, \citenamefont {Vaughan},
  \citenamefont {Wilke}, \citenamefont {Woo},\ and\ \citenamefont
  {Yutani}}]{Wickham2019}%
  \BibitemOpen
  \bibfield  {author} {\bibinfo {author} {\bibfnamefont {H.}~\bibnamefont
  {Wickham}}, \bibinfo {author} {\bibfnamefont {M.}~\bibnamefont {Averick}},
  \bibinfo {author} {\bibfnamefont {J.}~\bibnamefont {Bryan}}, \bibinfo
  {author} {\bibfnamefont {W.}~\bibnamefont {Chang}}, \bibinfo {author}
  {\bibfnamefont {L.}~\bibnamefont {McGowan}}, \bibinfo {author} {\bibfnamefont
  {R.}~\bibnamefont {Fran{\c{c}}ois}}, \bibinfo {author} {\bibfnamefont
  {G.}~\bibnamefont {Grolemund}}, \bibinfo {author} {\bibfnamefont
  {A.}~\bibnamefont {Hayes}}, \bibinfo {author} {\bibfnamefont
  {L.}~\bibnamefont {Henry}}, \bibinfo {author} {\bibfnamefont
  {J.}~\bibnamefont {Hester}}, \bibinfo {author} {\bibfnamefont
  {M.}~\bibnamefont {Kuhn}}, \bibinfo {author} {\bibfnamefont {T.}~\bibnamefont
  {Pedersen}}, \bibinfo {author} {\bibfnamefont {E.}~\bibnamefont {Miller}},
  \bibinfo {author} {\bibfnamefont {S.}~\bibnamefont {Bache}}, \bibinfo
  {author} {\bibfnamefont {K.}~\bibnamefont {M{\"{u}}ller}}, \bibinfo {author}
  {\bibfnamefont {J.}~\bibnamefont {Ooms}}, \bibinfo {author} {\bibfnamefont
  {D.}~\bibnamefont {Robinson}}, \bibinfo {author} {\bibfnamefont
  {D.}~\bibnamefont {Seidel}}, \bibinfo {author} {\bibfnamefont
  {V.}~\bibnamefont {Spinu}}, \bibinfo {author} {\bibfnamefont
  {K.}~\bibnamefont {Takahashi}}, \bibinfo {author} {\bibfnamefont
  {D.}~\bibnamefont {Vaughan}}, \bibinfo {author} {\bibfnamefont
  {C.}~\bibnamefont {Wilke}}, \bibinfo {author} {\bibfnamefont
  {K.}~\bibnamefont {Woo}},\ and\ \bibinfo {author} {\bibfnamefont
  {H.}~\bibnamefont {Yutani}},\ }\bibfield  {title} {\enquote {\bibinfo {title}
  {{Welcome to the Tidyverse}},}\ }\href {https://doi.org/10.21105/joss.01686}
  {\bibfield  {journal} {\bibinfo  {journal} {Journal of Open Source Software}\
  }\textbf {\bibinfo {volume} {4}},\ \bibinfo {pages} {1686} (\bibinfo {year}
  {2019})}\BibitemShut {NoStop}%
\bibitem [{\citenamefont {Salehi}\ \emph {et~al.}(2019)\citenamefont {Salehi},
  \citenamefont {Burkholder}, \citenamefont {Lepage}, \citenamefont {Pollock},\
  and\ \citenamefont {Wieman}}]{Salehi2019}%
  \BibitemOpen
  \bibfield  {author} {\bibinfo {author} {\bibfnamefont {S.}~\bibnamefont
  {Salehi}}, \bibinfo {author} {\bibfnamefont {E.}~\bibnamefont {Burkholder}},
  \bibinfo {author} {\bibfnamefont {G.~P.}\ \bibnamefont {Lepage}}, \bibinfo
  {author} {\bibfnamefont {S.}~\bibnamefont {Pollock}},\ and\ \bibinfo {author}
  {\bibfnamefont {C.}~\bibnamefont {Wieman}},\ }\bibfield  {title} {\enquote
  {\bibinfo {title} {{Demographic gaps or preparation gaps?: The large impact
  of incoming preparation on performance of students in introductory
  physics}},}\ }\href {https://doi.org/10.1103/PhysRevPhysEducRes.15.020114}
  {\bibfield  {journal} {\bibinfo  {journal} {Physical Review Physics Education
  Research}\ }\textbf {\bibinfo {volume} {15}},\ \bibinfo {pages} {20114}
  (\bibinfo {year} {2019})}\BibitemShut {NoStop}%
\bibitem [{\citenamefont {Jafari}\ and\ \citenamefont
  {Ansari-Pour}()}]{Jafari}%
  \BibitemOpen
  \bibfield  {author} {\bibinfo {author} {\bibfnamefont {M.}~\bibnamefont
  {Jafari}}\ and\ \bibinfo {author} {\bibfnamefont {N.}~\bibnamefont
  {Ansari-Pour}},\ }\bibfield  {title} {\enquote {\bibinfo {title} {{Why, When
  and How to Adjust Your P Values? Citation: Jafari M, Ansari-Pour N. Why, When
  and how to adjust your P values?}}}\ }\href
  {https://doi.org/10.22074/cellj.2019.5992} {\bibfield  {journal} {\bibinfo
  {journal} {Cell Journal(Yakhteh)}\ }\textbf {\bibinfo {volume} {20}},\
  \bibinfo {pages} {604--607}}\BibitemShut {NoStop}%
\bibitem [{\citenamefont {Benjamini}\ and\ \citenamefont
  {Hochberg}(1995)}]{Benjamini1995}%
  \BibitemOpen
  \bibfield  {author} {\bibinfo {author} {\bibfnamefont {Y.}~\bibnamefont
  {Benjamini}}\ and\ \bibinfo {author} {\bibfnamefont {Y.}~\bibnamefont
  {Hochberg}},\ }\bibfield  {title} {\enquote {\bibinfo {title} {{Controlling
  the False Discovery Rate: A Practical and Powerful Approach to Multiple
  Testing}},}\ }\href {https://doi.org/10.1111/J.2517-6161.1995.TB02031.X}
  {\bibfield  {journal} {\bibinfo  {journal} {Journal of the Royal Statistical
  Society: Series B (Methodological)}\ }\textbf {\bibinfo {volume} {57}},\
  \bibinfo {pages} {289--300} (\bibinfo {year} {1995})}\BibitemShut {NoStop}%
\bibitem [{\citenamefont {{U.S Department of Education}}\ and\ \citenamefont
  {{Office of Educational Technology}}(2017)}]{U.SDepartmentofEducation2017}%
  \BibitemOpen
  \bibfield  {author} {\bibinfo {author} {\bibnamefont {{U.S Department of
  Education}}}\ and\ \bibinfo {author} {\bibnamefont {{Office of Educational
  Technology}}},\ }\href {https://tech.ed.gov/files/2017/01/Higher-Ed-NETP.pdf}
  {\enquote {\bibinfo {title} {Reimagining the role of technology in higher
  education},}\ }\bibinfo {type} {Tech. Rep.}\ \bibinfo {number} {January}\
  (\bibinfo {year} {2017})\BibitemShut {NoStop}%
\bibitem [{\citenamefont {Maldonado-Mahauad}\ \emph {et~al.}(2018)\citenamefont
  {Maldonado-Mahauad}, \citenamefont {P{\'{e}}rez-Sanagust{\'{i}}n},
  \citenamefont {Kizilcec}, \citenamefont {Morales},\ and\ \citenamefont
  {Munoz-Gama}}]{Maldonado-Mahauad2018}%
  \BibitemOpen
  \bibfield  {author} {\bibinfo {author} {\bibfnamefont {J.}~\bibnamefont
  {Maldonado-Mahauad}}, \bibinfo {author} {\bibfnamefont {M.}~\bibnamefont
  {P{\'{e}}rez-Sanagust{\'{i}}n}}, \bibinfo {author} {\bibfnamefont {R.~F.}\
  \bibnamefont {Kizilcec}}, \bibinfo {author} {\bibfnamefont {N.}~\bibnamefont
  {Morales}},\ and\ \bibinfo {author} {\bibfnamefont {J.}~\bibnamefont
  {Munoz-Gama}},\ }\bibfield  {title} {\enquote {\bibinfo {title} {{Mining
  theory-based patterns from Big data: Identifying self-regulated learning
  strategies in Massive Open Online Courses}},}\ }\href
  {https://doi.org/10.1016/j.chb.2017.11.011} {\bibfield  {journal} {\bibinfo
  {journal} {Computers in Human Behavior}\ }\textbf {\bibinfo {volume} {80}},\
  \bibinfo {pages} {179--196} (\bibinfo {year} {2018})}\BibitemShut {NoStop}%
\bibitem [{\citenamefont {Saint}\ \emph {et~al.}(2021)\citenamefont {Saint},
  \citenamefont {Fan}, \citenamefont {Singh}, \citenamefont {Gasevic},\ and\
  \citenamefont {Pardo}}]{Saint2021}%
  \BibitemOpen
  \bibfield  {author} {\bibinfo {author} {\bibfnamefont {J.}~\bibnamefont
  {Saint}}, \bibinfo {author} {\bibfnamefont {Y.}~\bibnamefont {Fan}}, \bibinfo
  {author} {\bibfnamefont {S.}~\bibnamefont {Singh}}, \bibinfo {author}
  {\bibfnamefont {D.}~\bibnamefont {Gasevic}},\ and\ \bibinfo {author}
  {\bibfnamefont {A.}~\bibnamefont {Pardo}},\ }\enquote {\bibinfo {title}
  {Using process mining to analyse self-regulated learning: A systematic
  analysis of four algorithms},}\ in\ \href
  {https://doi.org/10.1145/3448139.3448171} {\emph {\bibinfo {booktitle}
  {LAK21: 11th International Learning Analytics and Knowledge Conference}}}\
  (\bibinfo  {publisher} {Association for Computing Machinery},\ \bibinfo
  {address} {New York, NY, USA},\ \bibinfo {year} {2021})\ p.\ \bibinfo {pages}
  {333–343}\BibitemShut {NoStop}%
\bibitem [{\citenamefont {Zimmerman}(2000)}]{Zimmerman2000}%
  \BibitemOpen
  \bibfield  {author} {\bibinfo {author} {\bibfnamefont {B.~J.}\ \bibnamefont
  {Zimmerman}},\ }\bibfield  {title} {\enquote {\bibinfo {title} {{Attaining
  self-regulation: A social cognitive perspective.}}}\ }in\ \href
  {https://doi.org/10.1016/B978-012109890-2/50031-7} {\emph {\bibinfo
  {booktitle} {Handbook of self-regulation.}}}\ (\bibinfo  {publisher}
  {Academic Press},\ \bibinfo {address} {San Diego},\ \bibinfo {year} {2000})\
  pp.\ \bibinfo {pages} {13--39}\BibitemShut {NoStop}%
\bibitem [{\citenamefont {Zimmerman}(2013)}]{Zimmerman2013}%
  \BibitemOpen
  \bibfield  {author} {\bibinfo {author} {\bibfnamefont {B.~J.}\ \bibnamefont
  {Zimmerman}},\ }\bibfield  {title} {\enquote {\bibinfo {title} {{From
  Cognitive Modeling to Self-Regulation: A Social Cognitive Career Path}},}\
  }\href {https://doi.org/10.1080/00461520.2013.794676} {\bibfield  {journal}
  {\bibinfo  {journal} {Educational Psychologist}\ }\textbf {\bibinfo {volume}
  {48}},\ \bibinfo {pages} {135--147} (\bibinfo {year} {2013})}\BibitemShut
  {NoStop}%
\end{thebibliography}%
\end{document}